\documentclass[conference]{IEEEtran}
\usepackage[T1]{fontenc}
\usepackage{cite}
\usepackage{booktabs}
\usepackage{array}
\usepackage{url}
\usepackage[hidelinks]{hyperref}

\begin{document}

\title{Assessing the Forensic Viability of Android Memory Analysis Across Production Builds: A Cross-Version Study of Security Hardening and Structure Preservation}

\author{
\IEEEauthorblockN{Jayasimha Nannapaneni}
\IEEEauthorblockA{Department of Electrical Engineering and Computer Science, Florida Institute of Technology}
\IEEEauthorblockN{Sneha Sudhakaran}
\IEEEauthorblockA{Department of Electrical Engineering and Computer Science, Florida Institute of Technology}
}

\maketitle

\begin{abstract}
Android memory forensics recovers evidence that never touches disk: decrypted messages, session credentials, and the live internal state of a running application. The tools that perform this recovery depend on debug symbols embedded in \texttt{libart.so}, the Android Runtime library, to locate data structures and interpret their layout. Across recent releases Google has stripped most of that information from the binaries that ship on consumer phones as part of a broader security hardening effort, yet no prior work has measured how far the stripping has progressed on the devices examiners actually encounter, or whether the memory architecture beneath the stripped surface still resembles what the forensics literature describes. This paper measures the gap between the unstripped development builds researchers use and the production builds that ship on real hardware, using binaries extracted from Pixel factory images across Android 8 through Android 15. Static symbols fell from 20,495 entries to zero, dynamic symbols dropped by roughly 60 percent, and source file references disappeared entirely. A compressed fallback section in the Android 15 binary restores thousands of function names but carries no structure layouts. Source code review and memory map comparisons across Android 8 and 15 show that the heap spaces, garbage collector infrastructure, and allocation bitmaps remain structurally intact, with visible changes limited to naming and arithmetic optimization. Live validation on a rooted, fully stripped Pixel 7 confirms that the runtime entry point is still locatable through the dynamic symbol table, and that structural offsets pulled from a version-matched development build resolve to valid pointers inside production memory. The results suggest that forensic tool compatibility with modern Android remains achievable, though it now takes considerably more manual effort than before.
\end{abstract}

\begin{IEEEkeywords}
Android forensics, memory forensics, ART runtime, symbol stripping, \texttt{libart.so}, digital forensics
\end{IEEEkeywords}

\section{Introduction}

Roughly three out of four smartphones in active use run Android \cite{statcounter2025}. That share is what makes memory forensics, the analysis of a device's volatile memory rather than its persistent storage, central to mobile investigations. A memory capture can hold artifacts that never reach disk at all: decrypted chat content, session keys, and the live internal state of whatever application was running when the phone was seized. A disk image cannot recover any of it; that data exists only briefly, in RAM.

Android does not hold still, though. Google ships a new major version every year, and each release can change how the runtime executes applications, manages memory, and, most relevant here, how much debugging information survives into the binaries that reach consumer devices. Over the last several releases Google has steadily stripped debug symbols from production binaries as part of a broader hardening strategy: those symbols tell a forensic tool what a function does and where a data structure lives, and removing them also raises the cost of reverse engineering the binary, since a full symbol table doubles as a blueprint for return-oriented and jump-oriented exploitation \cite{shacham2007}.

At the center of this problem is \texttt{libart.so}, the shared library implementing the Android Runtime (ART). Engineering and userdebug builds ship it with a complete symbol table, full DWARF debug information, and source file references, the version researchers typically work with. Consumer devices ship something else: a production build that has been through a stripping process designed to remove nearly all of it.

Most published Android memory forensics research was built and tested against that unstripped version, on binaries with full debug information and development builds where each structure still carries a name. No phone that reaches an examiner's desk runs that build, though. The most directly relevant prior work, Babangida et al. \cite{bappah2025}, tracked 34 internal ART structures across Android 9 through 14 and found that 73.2 percent of structure member offsets changed between versions, concluding that offset-based forensic methods face a mounting maintenance burden. That result is well supported by their data, but the entire analysis relied on engineering builds with full DWARF symbols: it measures how far structures drift from version to version while assuming the debug information needed to find those structures in the first place will always be there. That assumption does not hold on a real device.

That leaves two open questions, which this paper addresses directly.

\noindent\textbf{RQ1:} To what extent does Google's security hardening of production Android binaries limit the debug symbols available for memory forensics, and what forensic capability remains despite that hardening?

\noindent\textbf{RQ2:} Are the core memory management structures in ART architecturally preserved across Android versions despite this hardening, and does that preservation allow same-version forensic tool compatibility across build types?

Answering them required four things: quantifying symbol removal in production \texttt{libart.so} binaries across versions, examining the AOSP source and git history for evidence of structural change, comparing an unstripped and a production build of the same release to isolate the effect of stripping on its own, and validating the findings live on a rooted, fully stripped Pixel 7. What follows is the first concrete measurement of the forensic information gap on a real production device, evidence that the memory architecture beneath the stripped surface is intact, and a working procedure for recovering the ART Runtime object on a stripped Android 15 build from offsets carried over from a version-matched development build.

\section{Background and Related Work}

\subsection{ART and Android Memory Management}

Since Android 5.0, applications run on the Android Runtime rather than the older Dalvik VM \cite{google2017artdalvik}, compiling DEX bytecode into native code ahead of time, just in time, or through a mix of both depending on the release \cite{chartier2017}. ART gives each process its own heap, split into specialized regions, typically a main space, zygote space, and non-moving space, plus a separate large object space for allocations above roughly 12KB \cite{aligombe2019,sudhakaran2020}. Live and dead objects are tracked through mark and live bitmaps, modified pages through a card table, and collection coordination through allocation and live stacks.

The collector itself has changed more than once. Android 8 introduced the Concurrent Copying (CC) collector as the default, replacing the Concurrent Mark-Sweep collector that ran through Android 7 \cite{aligombe2019}, allocating within fixed-size regions on a structure called RegionSpace and evacuating regions once their live object count drops low enough. Android 13 later introduced a userfaultfd-based Concurrent Mark-Compact collector that removes CC's read barrier \cite{androiddev2022,aosp2023cmc}, which is the default by Android 15 on supported devices, with CC retained as a fallback.

Small object allocation under CC does not go through the older Runs-of-Slots Allocator (RosAlloc), an assumption in some earlier forensics literature, but through a bump-pointer allocator called RegionTLAB that gives each thread its own allocation buffer inside a RegionSpace region. RosAlloc remains in the codebase, but since Android 8 its role has narrowed to backing the non-moving space, where JNI-pinned objects live \cite{aosp2024gcdebug}. This matters for forensic technique: methods built around RosAlloc's deterministic first-available-slot behavior \cite{bhatia2018} cannot assume that behavior describes the allocator actually handling most allocations on a modern device. All of this, for a forensic tool, is reached through \texttt{libart.so}: whatever information that binary retains determines whether the tool can locate the runtime instance, walk the heap, and enumerate live objects.

\subsection{Android Userland Memory Forensics Tools}

Acquisition of Android memory is itself non-trivial. LiME \cite{sylve2012}, a loadable kernel module, remains the standard for kernel-level acquisition, but it requires root and a module compiled against the exact kernel on the target, often impractical on a locked-bootloader device with a recent kernel update \cite{stuttgen2014,spreitzenbarth2015}. Userland alternatives, reading memory through \texttt{/proc/[pid]/mem} on rooted devices or through dynamic instrumentation with Frida \cite{ravnas2024}, sacrifice full physical coverage but are far easier to deploy, and for ART-focused work they capture what actually matters.

Several tools built on this userland approach recover ART-managed objects specifically. DroidScraper \cite{aligombe2019}, presented at RAID 2019, locates the Runtime instance through a static symbol lookup in \texttt{libart.so} and traverses the heap to recover roughly 90 percent of live objects; it targeted Android 8, 8.1, and 9, and depends on the \texttt{nm} utility finding the mangled symbol \texttt{\_ZN3art7Runtime9instance\_E} in the static symbol table, an assumption that held for those versions but not for a stripped Android 15 binary. Timeliner \cite{bhatia2018} reconstructs the order in which a user opened different app screens from the spatial ordering that first-available-slot allocators tend to produce, though it still needs to locate Activity-related structures through the same kind of symbol information. AmpleDroid \cite{sudhakaran2020} extends DroidScraper to the Large Object Space, and RetroScope \cite{saltaformaggio2016} reconstructs prior app screens from recovered GUI elements. Sudhakaran et al. \cite{sudhakaran2022} separately found that garbage collection timing and process foreground state at acquisition significantly affect how complete the recovered data turns out to be. Each of these tools was built and tested on Android 8 or 9 binaries with full or near-full symbol tables, and none of them addressed what happens once that information is stripped away.

\subsection{Security Hardening and Symbol Stripping}

Google's hardening measures span kernel-level protections such as ASLR, stack canaries, and Control Flow Integrity \cite{google2017hardening,abadi2009,google2018cfi}, but the change most relevant here operates at the binary level. Development builds ship \texttt{libart.so} with a full \texttt{.symtab}, DWARF debug sections \cite{dwarf2017}, a \texttt{.strtab}, and source file references, and production builds strip nearly all of that away. Two things survive: the \texttt{.dynsym} section, since the runtime linker needs it, and a compressed \texttt{.gnu\_debugdata} section \cite{redhat2012}, also called MiniDebugInfo, which embeds a minimal ELF containing function-level symbols for crash reporting.

The APEX module system introduced in Android 10 \cite{google2019apex} adds a further wrinkle: ART is now packaged as an independently updatable module, and \texttt{libart.so} moved from /system/lib64/ to /apex/com.android.art/lib64/, meaning it can be updated on its own schedule, separately from the base operating system.

\subsection{Cross-Version Runtime Evolution}

The study most directly relevant to this paper is Babangida et al. \cite{bappah2025}, which tracked 34 ART data structures across Android 9 through 14 on four architectures, extracting DWARF symbols with a custom parser built on pyelftools \cite{bendersky2024}. Their DWARF symbol counts grew from roughly 15,700 on Android 9 to about 19,100 on Android 14, and across the 34 structures tracked, 73.2 percent of members had their offsets change in at least two versions, alongside a DWARF version bump at Android 13 and a compression shift from zlib to zstd. Their conclusion, that symbol-based forensics faces a growing maintenance burden, is well supported by their data, but every binary in that study carried full DWARF information: the reported counts describe what a developer compiling AOSP with debug flags sees, not what ships to a customer. Once the symbols are gone, the relevant question is no longer whether an offset shifted between versions, but whether the structure can be located at all. That is a different problem, and, this paper argues, a prior one to the problem they measured.

Android's build documentation defines three variants \cite{aosp2026building}: eng, for day-to-day development with full symbols and relaxed security; userdebug, a middle configuration for pre-release evaluation; and user, what ships to retail devices, hardened and stripped. The eng or userdebug binaries used to validate all of the tools discussed above are not what ships on a retail device, and none of that prior work was tested against the user variant, the one an investigator actually finds on a seized phone. This paper targets that gap directly, quantifying what production stripping removes and testing whether the architecture underneath remains navigable once it has.

\section{Methodology}

The investigation proceeded in four phases: quantifying symbol availability in production binaries, examining the AOSP source and git history for structural change, isolating the effect of stripping by comparing an unstripped and a production build of the same release, and validating the findings empirically on live device memory.

\subsection{Equipment and Environment}

The primary device was a Google Pixel 7 ("Panther") running Android 15, build BP1A.250505.005.B1, rooted with Magisk \cite{wu2024magisk} by patching the \texttt{init\_boot} partition rather than boot, a partition-layout change Android 13 introduced. Root access was needed to read process memory maps through \texttt{/proc} and to pull binaries out of APEX modules. The earlier endpoint used an Android 8.0 (Oreo, API 26) emulator configured in Android Studio on a Pixel 2 XL, x86\_64 system image; although the architecture differs from the ARM64 Pixel 7, ART's memory management code and naming conventions are architecture-independent, so heap layout does not vary by target. All binary analysis ran on a Debian workstation using standard ELF utilities, \texttt{nm}, \texttt{readelf}, \texttt{objcopy} (the aarch64-linux-gnu cross variant), file, and \texttt{objdump}, with no custom tooling required.

\subsection{Phase 1: Production Binary Symbol Analysis}

Two versions bracket the stripping timeline: Android 8.0, Pixel 2 XL ("Taimen"), build OPD1.170816.010, the earliest version running ART's Concurrent Copying collector and DroidScraper's original target, and Android 15, Pixel 7 ("Panther"), build BP1A.250505.005.B1, the current production build on the test device and the version at which Android begins its transition to 16KB memory pages. The Android 15 binary was pulled directly from the device with \texttt{adb pull /apex/com.android.art/lib64/libart.so}; the Android 8 binary came from Google's public factory image archive \cite{google2025factory}, extracted by converting the sparse \texttt{system.img} to a raw image with \texttt{simg2img}, mounting it, and, since ART ships inside an APEX module from Android 10 onward, unzipping \texttt{com.android.art.apex} and mounting the payload image before copying the library out of lib64/.

Each extracted binary went through four checks: file to classify it as stripped or not, \texttt{nm} to count static symbol table entries (zero on a stripped binary, since \texttt{.symtab} is gone), \texttt{nm -D} to count the dynamic symbols that survive stripping because the runtime linker needs them, and \texttt{readelf -S} to enumerate every ELF section present, including whether \texttt{.symtab}, \texttt{.strtab}, any \texttt{.debug\_*} section, or \texttt{.gnu\_debugdata} exists. The Android 15 binary also carried a \texttt{.gnu\_debugdata} section, extracted and decompressed separately with \texttt{objcopy -{}-dump-section} followed by \texttt{xz -d}, then inspected with \texttt{readelf -s} to characterize exactly what that fallback mechanism provides.

\subsection{Phase 2: Source Code and Git History Analysis}

This phase moved from compiled binaries to the AOSP source tree itself \cite{google2025aosp}, examining art/runtime/gc/\texttt{heap.h} and \texttt{heap.cc}, the RosAlloc allocator's \texttt{rosalloc.h} and \texttt{rosalloc.cc}, the page-size handling code in bionic, and the Runtime class definition in \texttt{runtime.h}. Git history for the RosAlloc allocator specifically was traced to identify what had changed between Android 8 and the current codebase, and each modified call site was checked for whether the change touched a data structure or allocation algorithm, as opposed to a purely arithmetic optimization.

\subsection{Phase 3: Unstripped Build vs. Production Build Comparison}

Compiling AOSP from source with the eng flag was attempted first but failed for lack of memory on the available hardware. Instead, an unstripped symbols package was pulled from Google's Continuous Integration infrastructure at ci.android.com \cite{google2025ci}: branch aosp-android-latest-release, target aosp\_cf\_arm64\_only\_phone-userdebug, build ID 14981173. This is a userdebug build, not an eng target, but the distinction does not matter here, since both retain full DWARF information, complete static symbol tables, and source file references, exactly what this phase measures. The package targets a Cuttlefish virtual device instead of a Pixel 7, but \texttt{libart.so} compiles from the same device-independent source tree regardless of target, a claim the high dynamic symbol overlap reported in Section~\ref{sec:unstrippedvsprod} supports directly.

The production \texttt{libart.so} from the Pixel 7 and the CI build's \texttt{libart.so} were compared with the same tools used in Phase 1, plus a dynamic symbol overlap check: extracting the sorted, unique dynamic symbol names from each binary with \texttt{nm -D} and comparing them with \texttt{comm} to count symbols shared between the two builds versus symbols unique to each. A high overlap would confirm that the two binaries share a common codebase, and that the difference between them comes down to stripping, not source divergence.

\subsection{Phase 4: Empirical Cross-Version Validation}

The final phase tested whether the architectural preservation suggested by Phases 2 and 3 held up under direct observation, through three activities: a memory map comparison, a baseline replication of the original DroidScraper recovery on Android 8, and a live recovery attempt on the stripped Pixel 7.

For the memory map comparison, process memory maps for \texttt{system\_server} (chosen because it is a long-running process exercising the full range of ART's memory management) were pulled from \texttt{/proc/[pid]/maps} on both the Android 8 emulator and the Pixel 7, filtered for dalvik- and rosalloc-related entries, and compared by hand against each other. An earlier attempt to acquire full physical memory with LiME on the emulator failed because the available goldfish kernel source corresponded to a newer branch (4.4-dev) than the one the API 26 emulator actually ran, so \texttt{/proc}-based extraction was used instead.

Before attempting anything on Android 15, the original DroidScraper procedure \cite{aligombe2019} was replicated on Android 8 using GDB to locate the Runtime instance symbol and read the Runtime pointer from live memory, confirming the baseline recovery chain functions as documented when full symbols are present.

The core validation used Frida \cite{ravnas2024} to repeat that recovery on the Pixel 7. Because \texttt{\_ZN3art7Runtime9instance\_E} is a \texttt{GLOBAL PROTECTED} export required for inter-library linking, it survives stripping and remains in the dynamic symbol table: Frida's \texttt{Process.getModuleByName()} supplied the base address of \texttt{libart.so} at runtime, the known symbol offset located the instance pointer, and reading that address (process ID 32359) yielded the Runtime object itself. With the object located, DWARF field offsets for \texttt{heap\_} and \texttt{thread\_list\_} were pulled from the unstripped CI build with \texttt{readelf -wi} and applied directly to it; walking one level further, the offset for \texttt{bump\_pointer\_space\_} was applied to the recovered Heap object. A matching internal layout would resolve every one of these reads to a valid pointer; a diverged layout would return garbage or unmapped addresses instead.

Three further complications specific to modern Android were documented along the way: pointers on the Pixel 7 carry Memory Tagging Extension (MTE) tags in their upper bits \cite{arm2023mte,google2023mte,serebryany2018} that must be stripped before an address can be used for traversal; SELinux policy blocks direct \texttt{/proc/pid/mem} reads even with root \cite{google2024selinux,smalley2013}, which is why the recovery used Frida's in-process injection instead of an out-of-process read; and the APEX relocation of \texttt{libart.so} changes where a tool needs to look for the binary on disk. The full command history for all four phases, including the Frida and GDB scripts, is maintained in a public repository for reproducibility.

\section{Results}

\subsection{Symbol Availability Across Production Builds}

Table 1 summarizes the symbol counts extracted from production \texttt{libart.so} binaries at each endpoint.

\begin{table}[t]
\caption{Symbol Counts in Production \texttt{libart.so} Binaries}
\label{tab:symbols}
\centering
\small
\begin{tabular}{@{}lcc@{}}
\toprule
\textbf{Metric} & \textbf{Android 8} & \textbf{Android 15} \\
\midrule
Static symbols (\texttt{nm}) & 20,495 & 0 \\
Dynamic symbols (\texttt{nm -D}) & 6,577 & 2,614 \\
\texttt{readelf -{}-syms} total & 27,395 & 2,618 \\
Binary status (\texttt{file}) & not stripped & stripped \\
Source file references & 3,594 & 0 \\
\texttt{.symtab} present & yes & no \\
\texttt{.strtab} present & yes & no \\
\texttt{.debug\_*} sections & 0 & 0 \\
\texttt{.gnu\_debugdata} present & no & yes \\
\bottomrule
\end{tabular}
\end{table}

Static symbols collapsed from 20,495 on Android 8 to zero on Android 15. Android 8's \texttt{libart.so} is classified by file as not stripped; Android 15's is classified as stripped. Dynamic symbols, which cannot be removed without breaking runtime linking, still dropped by roughly 60 percent, from 6,577 (\texttt{nm -D}) to 2,614 (2,618 by \texttt{readelf}'s count; the four-entry difference comes from how each tool handles null and special entries). Source file references disappeared entirely, from 3,594 down to zero, and neither production build examined carried any DWARF debug section. Because only these two endpoints were analyzed at the factory-image level, the exact version at which complete static stripping began cannot be pinned down from this data alone; other manufacturers may follow a different timeline.

Android 15's binary does retain a \texttt{.gnu\_debugdata} section. Decompressed, it yields 19,287 symbol entries, about seven times the 2,618 dynamic symbols in the main table, but every entry is a bare \texttt{FUNC} symbol: a name and an address, nothing else. There are no type definitions, no structure member offsets, and no source mappings. That is enough to label which function occupies a given address range for a crash report, but not enough to tell a forensic tool where the Heap or Thread fields sit inside a recovered Runtime object.

\subsection{Source Code Evolution}
\label{sec:sourcecode}

Git history for \texttt{rosalloc.cc} traces to a single commit, \texttt{dd98d26e9b} ("Optimize division by / modulo of gPageSize," authored by Richard Neill at ARM), as the primary change to that file between Android 8 and the current tree. The commit replaced direct division and modulo operations against the page size with bitwise equivalents, \texttt{DivideByPageSize()} and \texttt{ModuloPageSize()}, across 47 call sites, explicitly to support "legacy 4K, page size agnostic 4K and 16K" configurations. Every one of the 47 modified sites involved only the calculation method; no data structure was added, removed, or reorganized, and no allocation algorithm changed. The larger structural shift in this period was not to RosAlloc itself but to what sits above it: Android 8 introduced the Concurrent Copying collector and its RegionSpace/RegionTLAB allocator as the new default for the main heap, which is why RosAlloc's own file needed so little internal change: its role narrowed, but the allocator itself was largely left alone.

\subsection{Unstripped Build vs. Production Build}
\label{sec:unstrippedvsprod}

Table 2 lists the ELF sections present in each build.

\begin{table}[t]
\caption{ELF Sections Present in Unstripped (CI) versus Production (Pixel~7) \texttt{libart.so}, Android 15}
\label{tab:elfsections}
\centering
\small
\begin{tabular}{@{}lcc@{}}
\toprule
\textbf{ELF Section} & \textbf{Unstripped} & \textbf{Production} \\
\midrule
\texttt{.dynsym} & present & present \\
\texttt{.gnu.version} & present & present \\
\texttt{.symtab} & present & absent \\
\texttt{.strtab} & present & absent \\
\texttt{.debug\_info} & present & absent \\
\texttt{.debug\_line} & present & absent \\
\texttt{.debug\_abbrev} & present & absent \\
\texttt{.debug\_str} & present & absent \\
\texttt{.debug\_addr} & present & absent \\
\texttt{.debug\_rnglists} & present & absent \\
\texttt{.debug\_loclists} & present & absent \\
\texttt{.debug\_aranges} & present & absent \\
\texttt{.debug\_str\_offsets} & present & absent \\
\texttt{.debug\_line\_str} & present & absent \\
\texttt{.gnu\_debugdata} & absent & present \\
\bottomrule
\end{tabular}
\end{table}

The CI symbols build carries ten separate \texttt{.debug\_*} sections along with a full \texttt{.symtab} and \texttt{.strtab}. The production build from the Pixel 7 has none of these, and instead carries the single \texttt{.gnu\_debugdata} section the CI build lacks. In raw symbol counts, the unstripped build shows 63,027 static symbols via \texttt{nm} against zero for production: a complete removal.

To confirm that the two binaries actually share a codebase, and not just a matching version number, their dynamic symbol tables were compared directly. Of the unique dynamic symbols in each, 2,385 were shared, a 90.4 percent overlap relative to the CI build's unique dynamic symbol count. The symbols that did not overlap were traceable to device-specific differences between the Cuttlefish virtual target and the physical Pixel 7, not to any divergence in source code, which supports treating stripping as the only meaningful difference between the two binaries for the purposes of this comparison.

\subsection{Empirical Validation on the Pixel 7}

Table 3 maps memory structure naming between the two versions.

\begin{table*}[t]
\caption{Dalvik/RosAlloc Memory Structure Naming, Android 8 versus Android 15}
\label{tab:memnaming}
\centering
\small
\begin{tabular}{@{}p{0.46\textwidth}p{0.46\textwidth}@{}}
\toprule
\textbf{Android 8} & \textbf{Android 15} \\
\midrule
\texttt{/dev/ashmem/dalvik-main space} & \texttt{[anon:dalvik-main space]} \\
\texttt{/dev/ashmem/dalvik-zygote space} & \texttt{[anon:dalvik-zygote space]} \\
\texttt{/dev/ashmem/dalvik-non moving space} & \texttt{[anon:dalvik-non moving space]} \\
\texttt{dalvik-card table} & \texttt{dalvik-card table} \\
\texttt{dalvik-live stack} & \texttt{dalvik-live stack} \\
\texttt{dalvik-allocation stack} & \texttt{dalvik-allocation stack} \\
\texttt{dalvik-rosalloc page map} & \texttt{dalvik-linear-alloc page-status map} \\
\texttt{dalvik-allocspace main rosalloc} & \texttt{dalvik-allocspace non moving} \\
\bottomrule
\end{tabular}
\end{table*}

Filtered memory maps for \texttt{system\_server} contained 284 dalvik- and rosalloc-related entries on the Android 8 emulator against 121 on the Pixel 7, a 57 percent reduction that looks more severe than it is: Android 15 consolidates several numbered sub-structures, several separate mark-bitmap entries, for instance, into fewer combined entries; the underlying tracking mechanism is not removed, just consolidated. Each core concept present on Android 8 (main space, zygote space, non-moving space, card table, allocation and live stacks, bitmap tracking) is still present on Android 15. The visible changes are a naming shift from explicit /dev/ashmem/ prefixes to anonymous [anon:] mappings, and the reassignment already noted in Section~\ref{sec:sourcecode}: dalvik-allocspace main rosalloc on Android 8 becomes dalvik-allocspace non moving on Android 15, with dalvik-rosalloc page map correspondingly renamed to dalvik-linear-alloc page-status map. Structures such as the card table, live stack, and allocation stack kept identical names across both versions.

The Runtime instance symbol itself, \texttt{\_ZN3art7Runtime9instance\_E}, appears in the dynamic symbol table of both the unstripped Android 8 build and the fully stripped Android 15 production build, marked \texttt{GLOBAL PROTECTED} in both, meaning Google cannot remove it without breaking the runtime's own ability to link at startup. Replicating the original DroidScraper procedure on the Android 8 emulator located this symbol at offset \texttt{0x0070a980} through GDB, matching the original paper, and successfully read the Runtime pointer from live memory. Repeating the same recovery on the Pixel 7 with Frida found the \texttt{libart.so} base address at \texttt{0x7125386000}; adding the known offset, \texttt{0xc0d238}, located the instance pointer at \texttt{0x7125f93238}, and reading that address returned the Runtime object at \texttt{0xb400007227ddcc40} (the leading b4 byte is an MTE tag; the untagged address is \texttt{0x7227ddcc40}). The same symbol, seven major Android versions later and under full static stripping, still resolves.

Locating the Runtime object is only the entry point, though. It occupies 3,072 bytes containing pointers to the Heap, the ThreadList, the ClassLinker, and dozens of other structures, and without DWARF information there is no way to tell which bytes correspond to which field just by looking at a hex dump. To locate individual fields inside it, DWARF offsets extracted from the unstripped CI build, \texttt{heap\_} at offset \texttt{0x200} within Runtime and \texttt{thread\_list\_} at \texttt{0x248}, were applied directly to the live Runtime object recovered above, and the offset for \texttt{bump\_pointer\_space\_}, at \texttt{0x338} within the Heap object, was applied one level further in. Every one of these reads resolved to a valid ARM64 pointer: \texttt{heap\_} to \texttt{0xb4000071e7ddecf0}, \texttt{thread\_list\_} to \texttt{0xb400007257de2010}, and \texttt{bump\_pointer\_space\_} to \texttt{0xb400007297de6210}. The production binary's internal layout matches the unstripped build's layout exactly. Stripping removes the labels, not the structure. This result is specific to matched versions, however. Given Babangida et al.'s finding that most structure offsets shift across versions, there is no reason to expect offsets from a different Android release to resolve correctly against an Android 15 binary, and that cross-version transfer was not attempted here.

\section{Discussion}

\subsection{Answering the Research Questions}

RQ1 asked how far security hardening limits symbol availability and what capability survives it. The impact is severe but not absolute: production \texttt{libart.so} lost every static symbol and most source references between Android 8 and 15, yet what remains is enough to start from, not enough to finish with. The Runtime instance symbol survives because linking requires it, the remaining dynamic symbols provide a handful of additional landmarks, and MiniDebugInfo supplies thousands of function names with addresses but no structural information. An examiner working only from what the production binary provides can locate the Runtime object itself, but nothing past it: the binary contains no map of which byte range inside that object corresponds to which field.

RQ2 asked whether the underlying memory architecture survives that hardening, and whether the answer enables tool compatibility across build types. Yes to both, with one qualification. The heap spaces, bitmap tracking, card table, and allocation stacks present on Android 8 all carry over to Android 15, changed in name and in the RosAlloc reassignment but not in kind. The Frida-based validation is the strongest evidence here: DWARF offsets from an Android 15 unstripped build resolved correctly against a production Pixel 7's live memory for every field tested, meaning a production binary's internal layout is identical to the unstripped build of that same version. The qualification is that this was shown within a single version, not across versions. Babangida et al.'s finding that most structure offsets shift between releases stands; this paper adds that, within a given release, an examiner who obtains a version-matched unstripped build can use its DWARF data directly as an offset map for the corresponding production device.

\subsection{Implications}

For practitioners, the workflow this points to is straightforward even if more effortful than the tools built for Android 8 assumed: identify the device's exact Android version and build number, pull a matching unstripped build from ci.android.com or by compiling AOSP, extract DWARF offsets from its \texttt{libart.so}, and apply them to the seized device's memory image. That is more manual work than DroidScraper or Timeliner required, but it is a defined, repeatable procedure.

For tool developers, a tool can no longer assume \texttt{nm} returns anything useful against a production Android 15 binary. Recovery logic needs to fall back to the dynamic symbol table for the Runtime instance and accept externally supplied DWARF offsets for everything past it. MTE tagging is a separate concern: any tool doing pointer arithmetic on a modern device needs to strip the tag bits first, or every subsequent read fails.

For the research community, validating a technique only against unstripped or engineering builds is no longer sufficient. The gap between what a researcher compiles and what an examiner is handed, negligible before Android 15's stripping regime, is now wide enough that an untested technique cannot be assumed to work on a production binary. Babangida et al.'s cross-version offset instability is relevant here too: the DWARF cross-application technique demonstrated in this paper needs a version-matched build for every release it is applied to, a maintenance burden, though a tractable one given that Google publishes CI builds for each release.

\subsection{Limitations}

Several constraints bound how far these findings generalize. All production binary analysis used Pixel devices; manufacturers that ship modified ART builds, Samsung, OnePlus, and Xiaomi among them, were not examined, so whether the same stripping and structural preservation hold on non-Pixel hardware remains untested. The factory-image analysis covered only the two endpoints, Android 8 and 15, so the exact progression between them, including the precise point at which static stripping became total, is not characterized. The Android 8 memory map came from an x86\_64 emulator while the Android 15 map came from an ARM64 device; ART's memory management code is architecture-independent, but this mismatch means the comparison reflects both a version and a potential architecture difference, though none was observed. Only \texttt{system\_server} was examined; other processes could show different layouts, though \texttt{system\_server} was chosen because it exercises the full range of ART's memory features. The DWARF cross-application was validated only within a single version, not across versions, and while the Frida validation confirmed the Runtime, Heap, and BumpPointerSpace pointers are reachable, full object enumeration on the production device was not attempted end to end. Finally, LiME could not be deployed on the Android 8 emulator due to a kernel mismatch, so all analysis relied on userland access through \texttt{/proc} and Frida rather than a full physical acquisition; there is no specific reason to expect different results from a kernel-level capture, since the relevant structures live in userland memory regardless of method, but that assumption was not tested.

\section{Conclusion and Future Work}

Android 15 marks a real turning point for production binary stripping. Static symbols, debug information, and source file references are gone entirely from the binaries examiners actually receive, and any forensic tool built or tested before this shift now works in a fundamentally different environment. What has not changed is the architecture underneath. The structures present on Android 8 are all still there on Android 15, carrying new names but the same layout, and the Runtime instance symbol that any recovery chain starts from survives because the system cannot function without it. The measured gap, over 63,000 symbols in a development build against zero in production, is real, but nothing about it required changing how the runtime is built or laid out in memory. The structures are still where the source code says they should be, and a version-matched unstripped build can supply the map the device itself no longer does.

Several directions follow from this work. The DWARF cross-application demonstrated here was done entirely by hand. Automating the retrieval of a matching CI build and the extraction of its offsets would move the technique out of the research setting and into routine casework. Testing whether the same stripping and structural preservation hold on Samsung, OnePlus, Xiaomi, and other manufacturers' builds is an open question, since vendor modifications to ART could cut either way. Extending the Frida-based recovery beyond Runtime traversal to full object enumeration on a production device would close the gap between a structure being reachable and evidence actually being recovered. A more ambitious direction is version-independent structure discovery: if ART structures carry byte patterns or pointer relationships that hold steady across releases, scanning for those patterns might locate them without symbols or precomputed offsets at all, removing the need for a version-matched build. Finally, longitudinal tracking of future releases would show whether Google continues down this path, whether the \texttt{.gnu\_debugdata} fallback eventually disappears too, and whether the Runtime instance symbol remains exported or is made internal, since that symbol is the one reliable way into the runtime that this study identified.

\section*{Acknowledgment}
This work is based, in part, on research conducted as part of the first author's Master's thesis in the Department of Electrical Engineering and Computer Science at Florida Institute of Technology. A substantially expanded study building on this work is currently in preparation for journal submission.

\end{document}